\documentclass[a4paper,11pt]{article}
\pdfoutput=1 

\usepackage{jinstpub} 

\usepackage[T1]{fontenc} 

\usepackage[normalem]{ulem} 

\usepackage{datetime}

\newcommand{\Nucl}[2]{$\mathrm{^{#2}#1}$}

\title{\boldmath Cathode signal in a TPC directional detector: implementation and validation measuring the drift velocity}

 \author[a]{C.~Couturier,}
 \author[a,b]{Q.~Riffard,}
 \author[a]{N.~Sauzet,}
 \author[a]{O.~Guillaudin,}
 \author[a]{F.~Naraghi}
 \author[a]{and D.~Santos}
\affiliation[a]{LPSC, UGA, CNRS/IN2P3, Avenue des Martyrs, 38000, Grenoble, France}
\affiliation[b]{APC, Universit\'e Paris Diderot, CNRS/IN2P3, CEA/Irfu, Obs. de Paris, Sorbonne Paris Cit\'e, 75205 Paris, France}

\date{\today}
\emailAdd{ccouturi@lpsc.in2p3.fr}

\abstract{
	Low-pressure gaseous TPCs are well suited detectors to correlate  the directions of nuclear recoils
	to the galactic  Dark Matter (DM) halo. 
	Indeed, in addition to providing a measure of the energy deposition due to the elastic scattering of a DM particle on a nucleus in the target gas, they allow for the reconstruction of the track of the recoiling nucleus. 
	In order to exclude the background events originating from radioactive decays on the surfaces of the detector materials within the drift volume, efforts are ongoing to precisely localize the track nuclear recoil in the drift volume along the axis perpendicular to the cathode plane.
	We report here the implementation of the measure of the signal induced on the cathode by the motion of the primary electrons toward the anode in a MIMAC chamber. 
	As a validation, we performed an independent measurement of the drift velocity of the electrons in the considered gas mixture, correlating in time the cathode signal with the measure of the arrival times of the electrons on the anode.
}

\keywords{Dark Matter detectors; Micropattern gaseous detectors; Time projection chambers; Charge transport and multiplication in gas}

\begin{document}

\maketitle
\flushbottom

\section{Introduction -- fiducialisation}
Low-pressure gaseous TPCs are well suited detectors to correlate the directions of nuclear recoils 
to the galactic  Dark Matter (DM) halo. 
Indeed, in addition to providing a measure of the energy deposition due to the elastic scattering of DM particle on a nucleus in the target gas, they allow for the reconstruction of the track of this recoiling nucleus. 
	
Efforts are ongoing to precisely locate the track formation in the drift volume along the axis perpendicular to the cathode/anode planes (\textit{z} coordinate, with the convention that \mbox{$z = 0$} at the anode). In fact, the most prominent recoil background noise comes from radioactive decays originating from the surfaces of the detector materials, especially the daughters of the \Nucl{Rn}{222} and \Nucl{Rn}{220} nuclei. A way to identify this background is to exclude events coming from the detector surfaces within the drift volume.

Basically, a TPC equipped with a pixelated 2D readout provides a 2-dimensional image of the projection of the track of the recoiling nucleus. A fast sampling system can additionally scan the track in the \textit{z}-axis and provide a 3-dimensional image of the track; however, this does not give access to the \textit{absolute} position of the track along the \textit{z}-axis. 
Several methods have been implemented to retrieve the absolute $z$ coordinate in a low-pressure gas TPC. DRIFT~\cite{daw_e_2012} adds O$_2$ to a CS$_2$-based gas mix to produce
minority carriers~\cite{Battat20151} along with the negative ions.   
The measure of the difference of time arrivals of the different species (negative ions, minority carriers) on the anode/grid leads to the calculation of the position of the track in the drift volume. The drawback of this method is that it leads to a decrease in gain as primary electrons are recombined along the track.
An other method uses the diffusion of the electron cluster to provide an estimation of the \textit{z} coordinate: it relies on the fact the standard deviation of the charge dispersion varies as the square root of the location \textit{z} of the track.
This has been used on MIMAC data in order to estimate the \textit{z} coordinate of nuclear recoil tracks, and help identifying the radon progeny recoils~\cite{riffard_2015}.

A third method of retrieving the absolute $z$ coordinate in a low-pressure gas TPC is the use of the current induced on the cathode by the movement of the primary electrons in the drift space, referred to as the ``Cathode signal''. In the following section, we will introduce the Cathode signal and how it can be correlated to the absolute position of a recoil track. As a first step, we implemented the measure of this signal for $\alpha$ particles from an $^{241}$Am source: this will be described in the next section. The results, linking the absolute $z$ position to the cathode signal, allow an independent measurement of the electron drift velocity in the considered gas mixture.

\section{Cathode signal}

In the drift chamber, a recoil looses a fraction of its energy by ionizing the target gas along its path.
The primary electrons produced along the recoil track are collected by an electric field toward the anode.
Using the Shockley-Ramo theorem~\cite{shockley_currents_1938,ramo_currents_1939}, it can be shown \cite{recine_understanding_2014} that the movement of a particle of charge $q$ in the drift space between the two parallel electrodes (anode and cathode) induces a current on the cathode given by:

\begin{equation}\label{eq:signalcathode}
	I= - \frac{q}{V_Q}\vec{E}_{Q} \cdotp {\vec{v}_{q}} 
\end{equation}
where the \textit{weighting field} $\vec{E}_{Q}$ is an hypothetical electric field obtained in the same geometry, with an electric potential of \mbox{$V_Q$ = 1V} on the cathode and all other electrodes grounded, and $\vec{v}_q$ is the velocity of the moving charge.

If we discard the side effects (infinite parallel electrodes), the weighting field can be written as $\vec{E}_{Q} = \frac{V_Q}{D} \vec{e_z}$, with $D$ the distance between the two parallel electrodes. Considering a primary electron going away from the cathode, and assuming that it reaches rapidly its maximal drift velocity, the velocity $\vec{v}_{q}$ can be written as $\vec{v}_{q}(z) = {v_{\rm drift}} \vec{e_z}$.

Thus, in the case of $N_e$ individual primary electrons with charge $-e$, the total current induced on the cathode is:
\begin{equation}\label{eq:signalcathode2}
	I= N_e \, e \, \frac{v_{\rm drift}}{D}
\end{equation}

Unlike in~\cite{recine_understanding_2014}, we do not work on the shape and duration of the  signal induced on the cathode. Instead, we are only interested in the time correlation of the cathode signal with respect to the avalanche at the anode level.
Indeed, this signal on the cathode starts as soon as the primary ionization electrons start moving, assuming that they rapidly reach  their maximal drift velocity $v_{\rm drift}$.
On the anode side, two components could contribute to the current induced on the grid: 
\begin{itemize}
    \item the signal induced in the grid by the motion of the primary electrons: indeed, the primary ionization electron signal shows up in both cathode and grid electrodes at the same time. However, the contribution of these primary electrons moving to the anode is too fast and is washed by the rise time of the preamplifier.

    \item the motion of ions produced during the avalanche going upwards to the grid: this is by far the main contribution. It starts once the primary electrons reach the anode.
\end{itemize}

The time $\Delta t$ between the two signals thus corresponds to the transit time of the primary electrons from their production to the grid.
Therefore it is possible to calculate the \textit{z} coordinate by measuring the time difference between the cathode signal and the signal at the grid, assuming we know the electron drift velocity $v_{\rm drift}$ in the gas mixture as a function of the position\footnote{For the actual MIMAC setup, though we do not work with infinite electrode plates, a field cage is enforcing a constant electric field in the drift volume: the drift velocity is constant with respect to the position \textit{z}.}.
Conversely, using a ionization source located in a known \textit{z} we can measure the electron drift velocity $v_{\rm  drift}$ with the time difference between the two signals. This is what will be demonstrated in the following sections.

\section{Experimental setup}

\begin{figure}[tbph]
	\centering
	
	\vspace{-3mm}
	\includegraphics[width=0.55\linewidth]{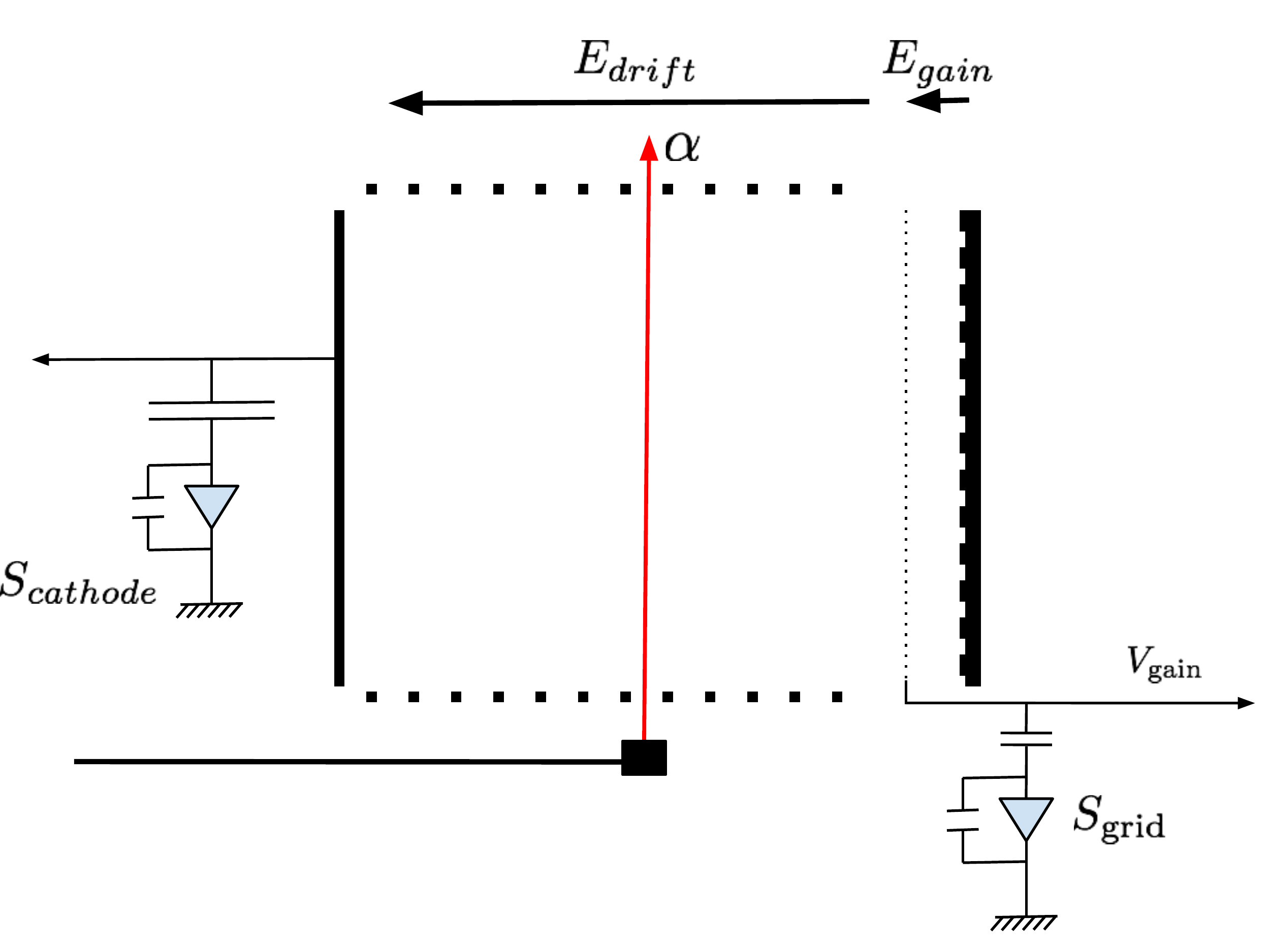}
	\caption{Scheme of the experimental setup: the collimated source provides a beam of $\alpha$ particles parallel to the cathode plane~\cite{riffard:tel-01258830}. }
	\label{fig:schema}	
\end{figure}

A MIMAC (MIcro-tpc MAtrix of Chambers) bi-chamber module~\cite{Santos2013} has been modified to allow an $^{241}$Am $\alpha$ source attached to a rod to be moved along the $z$ axis, \textit{i.e.} perpendicular to the cathode plane, without opening the chamber.
The source delivers a collimated ($\Delta\theta<0.01^\circ$) beam of $\alpha$'s oriented parallel to the cathode plane as shown in Figure~\ref{fig:schema}, with kinetic energies of 5.37\,MeV; a SRIM simulation shows that each $\alpha$ releases an ionization energy of the order of $\sim$MeV in the volume.
The drift chamber was filled with the following gas mixture: $70\%$ $\textrm{C}\textrm{F}_4$+$28\%$ $\textrm{C}\textrm{H}\textrm{F}_3$+$2\%$ $\textrm{C}_4\textrm{H}_{10}$, at a pressure of 50\,mbar.
The primary electrons produced by ionization along the $\alpha$ tracks are collected by an electric field of 154 V/cm toward the anode.
These electrons are amplified at the anode level by a 256\,$\mu$m-gap  Micromegas
with a 10.8\,cm\,$\times$\,10.8\,cm total area.
The signal on the grid of the Micromegas, coming from the avalanche in the 256\,$\mu$m-gap, is measured by a first channel ("anode signal").
An additional channel records the signal induced on the cathode by the \textit{movement} of the primary electrons  ("cathode signal")\footnote{Characteristics of the cathode preamplifier: Gain: $\sim$1\,V/pC, Rise time: 60\,ns, Power supply: $\pm$\,12V}.
A coincidence allows for measuring the time interval between the two signals.

The triggering signal is usually chosen as the less frequent signal: here the anode signal\footnote{Indeed, the surface of the anode grid is smaller than the surface of the cathode.}.
The signals $\Delta$TAC recorded by the Time Amplitude Converter corresponds to the time difference between the cathode signal, delayed (fixed duration), and the anode signal: $\Delta$TAC = ${\rm Delay} - \Delta t$, with $\Delta t$ the transit time of the primary electrons toward the anode, assuming the signal on the cathode is induced quasi-instantaneously.
The delay was chosen to make sure all the anode signals are measured before their corresponding cathode signal. The exact value of the total delay is not known, as it depends on electronic delays (\textit{e.g.} arising by difference of wire lengths).

In the following section we will analyze the measurements of the time differences between the signal from the anode and the signal from the cathode.

\section{Results and discussion}

\begin{figure}[tbph]
	\centering
	\includegraphics[width=0.59\linewidth]{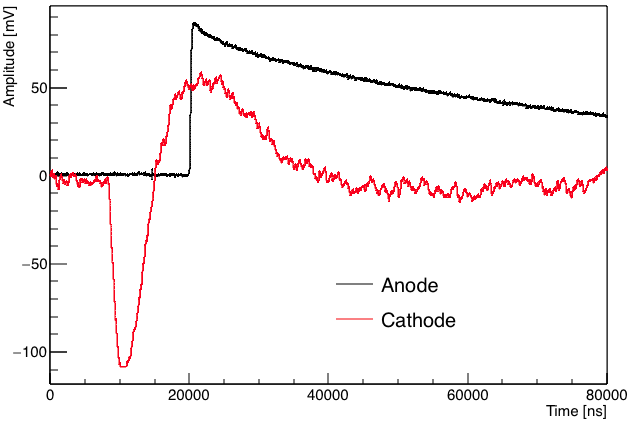}
	\includegraphics[width=0.39\linewidth]{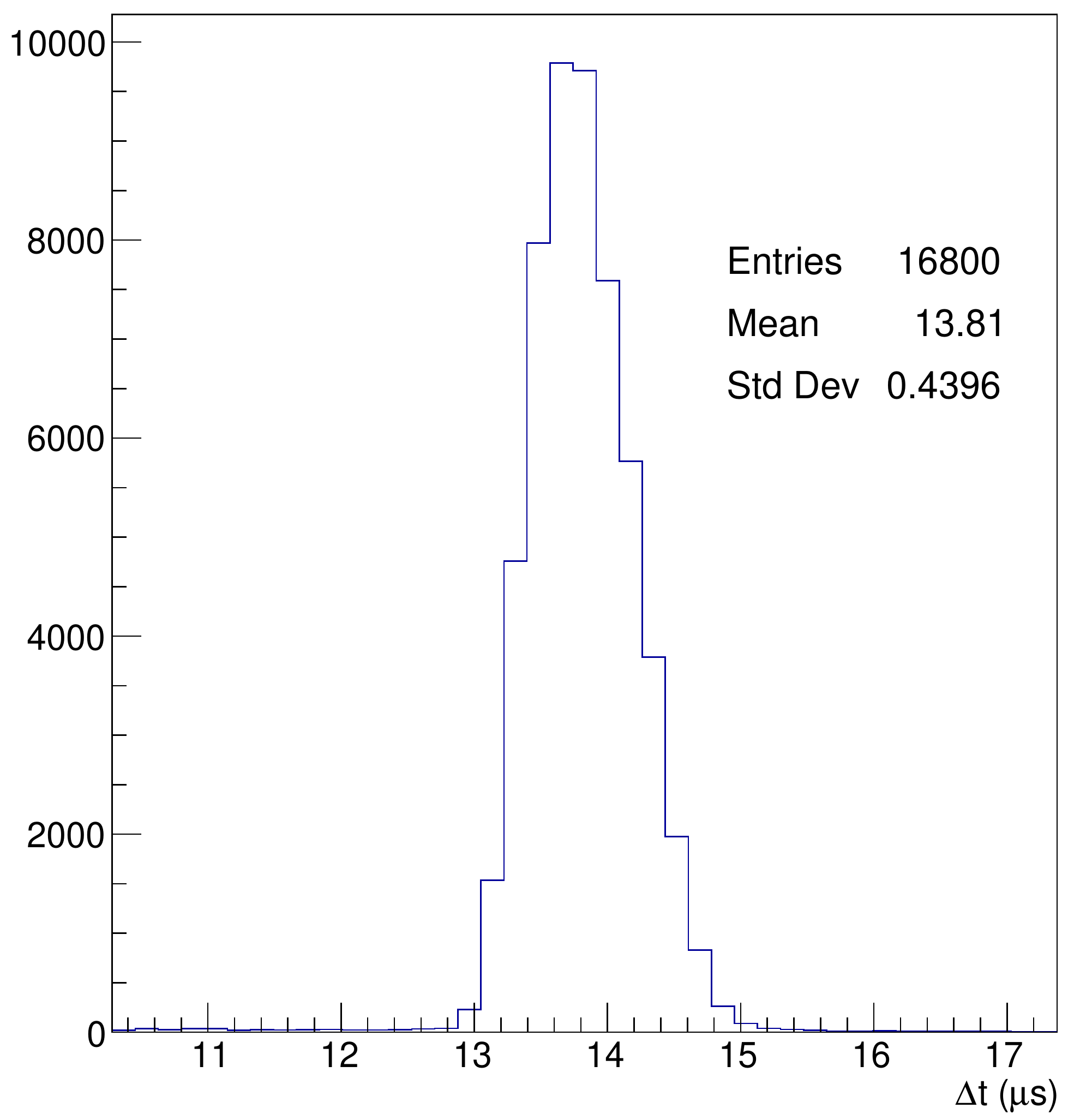}
	\caption{Left: amplitudes of the signals from the cathode preamplifier and from the grid preamplifier as shown on the oscilloscope. Note that here, the cathode signal is not delayed, and is detected before the anode signal.
Right: distribution of the $\Delta$TAC signals (with delayed cathode signal) with the $\alpha$ source located at 140\,mm of the anode. }
\label{fig:oscillo2}
\end{figure}

Figure \ref{fig:oscillo2} (left) shows the amplitudes of both preamplifier signals -- from the cathode and from the anode -- for one event (no delay here). 
The $\Delta$TAC is recorded as a digitized value (ADC units). 
Figure \ref{fig:oscillo2} (right) shows the  distribution of the $\Delta$TAC signals with the $\alpha$ source located at 140\,mm of the anode.
The conversion factor (TAC-to-$\mu$s) has been precisely obtained  with a dedicated measurement by a pulse generator to obtain the actual physical time difference: a$_{\rm {TAC-to-\mu}s} = 181.2$\,ADC/$4\mu$s.

\begin{figure}
	\centering
	\includegraphics[width=0.7\linewidth]{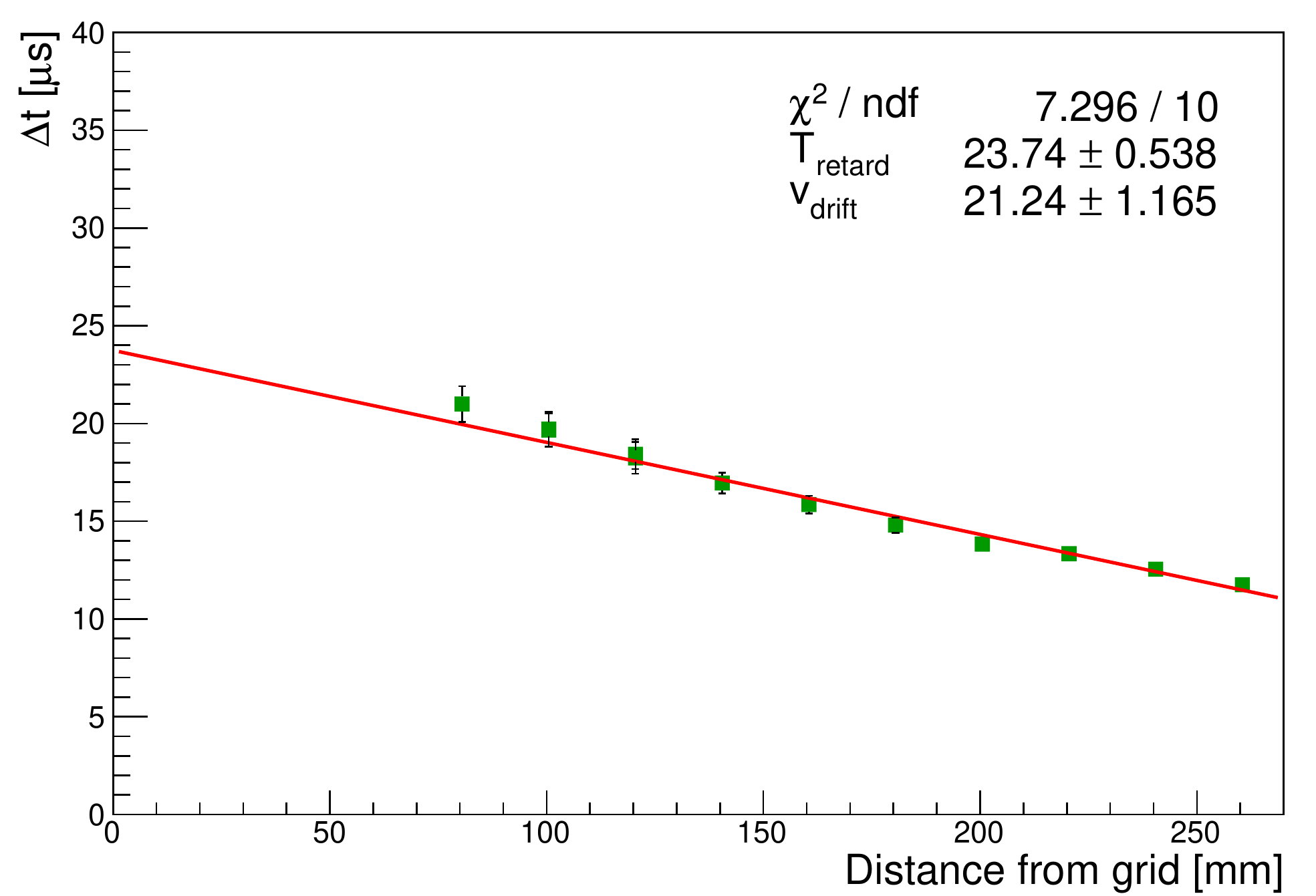}
	\caption{Measure of the time differences ($\Delta$TAC) between the grid signal and the delayed cathode signal  as a function of the distance of the $\alpha$ source from the anode (green points) ; only the statistical erors are displayed. A linear fit of these points is superimposed in red and provides the values of the drift velocity and the additional delay.}
	\label{fig:Start_Grid}
\end{figure}

We recorded the time differences $\Delta$TAC between the anode signal and the delayed cathode signal for several known positions of the alpha source.
This allows to plot the $\Delta$TAC values as a function of the distance of the $\alpha$ source to the anode, as shown in Figure \ref{fig:Start_Grid}. 
It presents a linear relationship between the time differences and the absolute \textit{z} positions of the parallel $\alpha$ tracks. This relation can be used to determine the (first unknown) $z$ absolute position from the (measured) $\Delta$TAC signal. It is still hard with current setup to measure the signal induced on the cathode (or equivalently the time differences) for charges moving at 20-25 cm of the cathode. One way of tackling this would be to use a larger cathode to increase the amplitude of the recorded induced signal.

The X-axis error bars are plotted and correspond to the reading error on the distance of the source to the anode (1\,mm).
The Y-axis error bars include the statistical errors on the $\Delta$TAC measurements (standard deviation), and the incertainty on the measure of the $\Delta$TAC position ($\sim 0.1\,\mu$s).
A linear fit of these points provides the values of the drift velocity and the time delay added:
\begin{itemize}
	\item total delay T$_{\rm delay} = (23.7 \pm 0.5) \mu$s ; it includes the additional delay as well as the electronic delays.
	\item drift velocity $v_{\rm drift} = (21.2\pm 1.2) \mu$m/ns.
\end{itemize}

The uncertainties on determining the calibration coefficient (~0.2\,ADC/$\mu$s) entails a systematic error of 0.04\,$\mu$m/ns on the value of the drift velocity.
The final value of the drift velocity reads:
$$ v_{\rm drift} = (21.2\pm 1.2 {\rm \tiny (stat)} \pm 0.04 {\rm \tiny (sys)}) \mu{\rm m/ns}.$$
It differs slightly from the value obtained using the MAGBOLTZ software~\cite{Biagi1999234}: 
$$v_{\rm drift} = (24.1\pm 0.02) \mu{\rm m/ns}.$$
The difference can be explained by the fact we do not work with the same conditions as described in the simulation: the pressure of the gas varies with the temperature of the room ($\sim$1\,mbar), the gas mix contains impurities, and the electric field in the drift volume can contain some amount of inhomogeneities. The measurement of the drift velocity performed with the method described here corresponds to a setup with real experimental conditions, which are not fully taken into account when performing the MAGBOLTZ simulation. 

An other \textit{in situ} measurement of the electron drift velocity\footnote{Other methods to measure the electron drift velocity in low pressure gas TPCs exist, e.g. using a laser~\cite{COLAS2002215}. For high pressure gas TPCs, one method consists in measuring the time difference between the primary scintillation produced by the radiation in the gas and the time of arrival of the ionization electrons to the ``anode''~\cite{Alvarez2013}; however, to our knowledge, this method has not proven successful to measure the electron drift velocity in low pressure gas TPCs.} was previously performed in a MIMAC chamber~\cite{billard_situ_2014} with a similar gas mixture ($70\%$ $\textrm{C}\textrm{F}_4$+$30\%$ $\textrm{C}\textrm{H}\textrm{F}_3$), at a pressure of 50\,mbar, for several values of the drift electric field. A different approach was applied, using $\alpha$ particles crossing the entire drift volume between the cathode and the anode, and using fast sampling of the pixelated anode signal.
Fig. 11 in this reference presents the measurements of the drift velocity in this similar gas mix at a pressure of 50\,mbar, with various reduced drift fields $E/N$ ranging from 14\,V.cm$^2$ to 21\,V.cm$^2$.
The closest data point in their measurements closest to our setup\footnote{We used a drift field of $E = 154$\,V/cm at a pressure of 50\,mbar, which corresponds to a reduced field of \mbox{$E/N = 12.5 \cdot 10^{17}$\,V.cm$^2$}.
(For an ideal gas at 1\,atm, \mbox{$N = 2.5 \cdot 10^{25} \rm{\,m}^{-3} = 2.5 \cdot 10^{19}$\,cm$^{-3}$}. At 50\,mbar, \mbox{$N = 2.5 \cdot 10^{19} \cdot \frac{50}{1013.25} = 1.23 \cdot 10^{18}$\,cm$^{-3}$)}.} is \mbox{$E/N = 14$\,V.cm$^2$}, for which they measured a drift velocity of 24.5\,$\mu$m/ns, slightly below the value of 27.4\,$\mu$m/ns computed with MAGBOLTZ.

\section{Conclusion}
We demonstrated the ability of a TPC experiment like MIMAC to achieve the measurement of the signal induced by the primary ionization electrons on the cathode, using an alpha source moved along the \textit{z}-axis. In particular, we have shown that we are able to make a measurement of the electron drift velocity when knowing the location of the source and measuring the time difference between the grid and cathode signals. Conversely, assuming a known electron drift velocity, we are now able to locate a recoil track along the \textit{z}-axis. 

Next step is to show that the  measurement of the cathode signal is feasible not only for MeV $\alpha$ particles, but also for lower energy nuclear recoils, in the Dark Matter search range ($\sim 10$\,keV).

\bibliographystyle{JHEP}
\bibliography{cathode}

\end{document}